\documentclass[conference]{IEEEtran}
\ifCLASSINFOpdf
\else
   \usepackage[dvips]{graphicx}
   \newcommand{\bysame}{%
    \leavevmode\hbox to 3em{\hrulefill}\,}
\fi
\begin{document}
%
\title{Simultaneous Code/Error-Trellis Reduction for Convolutional Codes Using Shifted Code/Error-Subsequences}

\author{\IEEEauthorblockN{Masato Tajima}
\IEEEauthorblockA{Graduate School of Sci. and Eng.\\
University of Toyama\\
3190 Gofuku, Toyama 930-8555, Japan\\
Email: tajima@eng.u-toyama.ac.jp}
\and
\IEEEauthorblockN{Koji Okino}
\IEEEauthorblockA{Information Technology Center\\
University of Toyama\\
3190 Gofuku, Toyama 930-8555, Japan\\
Email: okino@itc.u-toyama.ac.jp}
\and
\IEEEauthorblockN{Takashi Miyagoshi}
\IEEEauthorblockA{Graduate School of Sci. and Eng.\\
University of Toyama\\
3190 Gofuku, Toyama 930-8555, Japan\\
Email: miyagosi@eng.u-toyama.ac.jp}}


%


\maketitle

\begin{abstract}
In this paper, we show that the code-trellis and the error-trellis for a convolutional code can be reduced simultaneously, if reduction is possible. Assume that the error-trellis can be reduced using shifted error-subsequences. In this case, if the identical shifts occur in the subsequences of each code path, then the code-trellis can also be reduced. First, we obtain pairs of transformations which generate the identical shifts both in the subsequences of the code-path and in those of the error-path. Next, by applying these transformations to the generator matrix and the parity-check matrix, we show that reduction of these matrices is accomplished simultaneously, if it is possible. Moreover, it is shown that the two associated trellises are also reduced simultaneously.
\end{abstract}


%
\IEEEpeerreviewmaketitle

\section{Introduction}
In this paper, we always assume that the underlying field is $F=\mbox{GF}(2)$. Let $G(D)$ and $H(D)$ be the generator matrix and the parity-check matrix of an $(n, n-m)$ convolutional code $C$, respectively. Ariel and Snyders [1] presented a construction of error-trellises based on the scalar check matrix derived from $H(D)$. They showed that when some ($j$th) ``column'' of $H(D)$ has a factor $D^l$, there is a possibility that state-space reduction can be realized. Being motivated by their work, we also examined the same case. The time-$k$ error $\mbox{\boldmath $e$}_k=(e_k^{(1)}, \cdots, e_k^{(n)})$ and syndrome $\mbox{\boldmath $\zeta$}_k=(\zeta_k^{(1)}, \cdots, \zeta_k^{(m)})$ are connected with the relation $\mbox{\boldmath $\zeta$}_k=\mbox{\boldmath $e$}_kH^T(D)$ ($T$ means transpose). From this relation, we noticed [9] that the transformation $e_k^{(j)}\rightarrow D^le_k^{(j)}=e_{k-l}^{(j)}$ is equivalent to dividing the $j$th column of $H(D)$ by $D^l$. That is, reduction can be realized by shifting the ``subsequence'' $\{e_k^{(j)}\}$ of the original error-path {\boldmath $e$}. It is stated [1] that their construction can be used also to obtain code-trellises. However, it is not described in the paper. On the other hand, our construction is based on an equivalent modification of the relation $\mbox{\boldmath $\zeta$}_k=\mbox{\boldmath $e$}_kH^T(D)$. Hence, our method can be directly extended to code-trellises. That is, in the case of code-trellises, the construction is based on the relation $\mbox{\boldmath $y$}_k=\mbox{\boldmath $u$}_kG(D)$ and its equivalent modifications, where $\mbox{\boldmath $u$}_k$ and $\mbox{\boldmath $y$}_k$ are the time-$k$ information and code symbols, respectively. Note that there exists a one-to-one correspondence between the code-paths in a code-trellis and the error-paths in the corresponding error-trellis. Accordingly, it is reasonable to think that the two trellises can be reduced simultaneously, if reduction is possible. Here, consider the situation that the identical shifts occur both in the components of $\mbox{\boldmath $y$}_k$ and in those of $\mbox{\boldmath $e$}_k$. In this case, if one trellis is reduced, then the other trellis should be equally reduced. In this paper, based on this idea, we discuss the simultaneous reduction of a code-trellis and the corresponding error-trellis. First, we obtain the general transformations which generate the identical shifts both in the subsequences of {\boldmath $y$} and in those of {\boldmath $e$}. Next, we show that these transformations preserve the relation that {\it one is a generator matrix and the other is the corresponding parity-check matrix}. (In this paper, we call this relation the ``{\it GH Relation}'' and if $G(D)$ and $H(D)$ have this relation, then it is denoted as $G(D)\Leftrightarrow H(D)$). Using this property, it is shown that $G(D)$ and $H(D)$ are reduced simultaneously, if reduction is possible. Moreover, it is shown that the corresponding two trellises are also reduced simultaneously. These results again imply that a code/error-trellis construction using shifted code/error-subsequences is very effective.

\section{Trellis construction using shifted path-subsequences}
\subsection{Error-trellis construction using shifted error-subsequences}
Let $H(D)$ be the parity-check matrix for an $(n, n-m)$ convolutional code $C$. Consider the error-trellis based on the syndrome former $H^T(D)$. In this case, the adjoint-obvious realization of $H^T(D)$ is assumed unless otherwise specified. Assume that the $j$th column of $H(D)$ has the form
\begin{equation}
\left(
\begin{array}{cccc}
D^{l_j}h_{1j}'(D) &D^{l_j}h_{2j}'(D) &\ldots &D^{l_j}h_{mj}'(D)
\end{array}
\right)^T ,
\end{equation}
where $l_j\geq 1$. Let $H'(D)$ be the modified version of $H(D)$ with the $j$th column being replaced by
\begin{equation}
\left(
\begin{array}{cccc}
h_{1j}'(D) &h_{2j}'(D) &\ldots &h_{mj}'(D)
\end{array}
\right)^T .
\end{equation}
Also, let $\mbox{\boldmath $e$}_k'\stackrel{\triangle}{=}(e_k^{(1)}, \cdots, e_k'^{(j)}, \cdots, e_k^{(n)})$, where $e_k'^{(j)}\stackrel{\triangle}{=}D^{l_j}e_k^{(j)}=e_{k-l_j}^{(j)}$. Then we have
\begin{equation}
\mbox{\boldmath $\zeta$}_k=\mbox{\boldmath $e$}_k'H'^{T}(D) .
\end{equation}
Hence, in the case where the $j$th column of $H(D)$ has a factor $D^{l_j}$, there is a possibility that an error-trellis with reduced number of states can be constructed by shifting the $j$th error-subsequence by $l_j$ time units [9]. Assume that the corresponding code-trellis is terminated in the all-zero state at $t=N$. Then $e_k'^{(j)}=e_{k-l_j}^{(j)}$ is modified as $e_k'^{(j)}=e_{<k-l_j>}^{(j)}$, where $<t>$ denotes $t\:\mbox{mod}\:(N+l_j)$ (i.e., ``cyclic shift'').

\subsection{Error-trellis construction using backward-shifted error-subsequences}
The construction using shifted error-subsequences is further extended [9], [10]. That is, a reduced error-trellis can be equally constructed using ``backward-shifted'' error-subsequences. Consider the transformation $e_k^{(j)}\rightarrow D^{-l_j}e_k^{(j)}=e_{k+l_j}^{(j)}$. We see that this is equivalent to ``multiplying'' the $j$th column of $H(D)$ by $D^{l_j}$. Let $H'(D)$ be the parity-check matrix after modification. If $H'(D)$ is reduced to an equivalent $H''(D)$ with overall constraint length less than that of $H(D)$, then reduction can be realized. We remark that the power $l_j$ of $D$ has to be determined properly for each $j$. For the purpose, we can use the {\it reciprocal dual encoder} [6] $\tilde H(D)$ associated with $H(D)$.
\par
{\it Example 1 ([9]):} Consider the canonical parity-check matrix
\begin{equation}
H_1(D)=\left(
\begin{array}{ccc}
D^2& D^2 & 1 \\
1& 1+D+D^2 & 0 
\end{array}
\right) .
\end{equation}
Since all the columns of $H_1(D)$ are delay free, any further reduction seems to be impossible. In fact, it follows from Theorem 1 of [1] that the dimension $d_1$ of the state space of the error-trellis based on $H_1^T(D)$ is $4$. However, a corresponding generator matrix is given by $G_1(D)=(1+D+D^2, 1, D^3+D^4)$. Observe that the third ``column'' of $G_1(D)$ has a factor $D^2$. ({\it Remark:} It suffices to divide the third column by $D^2$ in order to obtain a reduced code-trellis.) This fact implies that a reduced error-trellis can be constructed [1], [9]. Then consider the reciprocal dual encoder
\begin{equation}
\tilde H_1(D)=\left(
\begin{array}{ccc}
1& 1 & D^2 \\
D^2& 1+D+D^2 & 0 
\end{array}
\right) .
\end{equation}
Note that the third column of $\tilde H_1(D)$ has a factor $D^2$. Accordingly, dividing the third column of $\tilde H_1(D)$ by $D^2$, we can construct an error-trellis with $4$ states (i.e., $\tilde d_1=2$) [1], [9]. Here, notice that each error-path in the error-trellis based on $H_1^T(D)$ can be represented in time-reversed order using the error-trellis based on $\tilde H_1^T(D)$. Hence, a factor $D^2$ in the column of $\tilde H_1(D)$ corresponds to backward-shifting by two time units (i.e., $D^{-2}$) in terms of the original $H_1(D)$. Hence, multiply the third column $H_1(D)$ by $D^2$. Then we have
\begin{equation}
H_1'(D)=\left(
\begin{array}{ccc}
D^2& D^2 & D^2 \\
1& 1+D+D^2 & 0 
\end{array}
\right) .
\end{equation}
We see that this matrix can be reduced to an equivalent canonical parity-check matrix
\begin{equation}
H_1''(D)=\left(
\begin{array}{ccc}
1& 1 & 1 \\
1& 1+D+D^2 & 0 
\end{array}
\right)
\end{equation}
by dividing the first ``row'' by $D^2$. Hence, the dimension $d_1$ can be reduced to $2$.

\subsection{Code-trellis construction using shifted code-subsequences}
Note that the relation $\mbox{\boldmath $y$}_k=\mbox{\boldmath $u$}_kG(D)$ holds with respect to a generator matrix $G(D)$, where $\mbox{\boldmath $u$}_k=(u_k^{(1)}, \cdots, u_k^{(n-m)})$ and $\mbox{\boldmath $y$}_k=(y_k^{(1)}, \cdots, y_k^{(n)})$ are the time-$k$ information and code symbols, respectively. In the same way as for $H(D)$, by dividing the $j$th column of $G(D)$ by $D^{l_j}$ or by multiplying the $j$th column of $G(D)$ by $D^{l_j}$, reduction of $G(D)$ can be realized. We see that the former corresponds to the backward-shift $y_k^{(j)}\rightarrow y_{k+l_j}^{(j)}$, whereas the latter corresponds to the forward-shift $y_k^{(j)}\rightarrow y_{k-l_j}^{(j)}$. Note that the shift directions are reversed compared to $H(D)$.

\section{Transformations generating the identical shifts both in {\boldmath $y$} and in {\boldmath $e$}}
\subsection{General case}
Consider the transformations which generate the identical shifts both in the components of $\mbox{\boldmath $y$}_k$ and in those of $\mbox{\boldmath $e$}_k$. Now, assume that the relation $G(D)\Leftrightarrow H(D)$ holds. Consider a pair of transformations:
\begin{enumerate}
\item divide the $j$th column of $G(D)$ by $D^{l_j^{(d)}}$ and multiply the same column by $D^{l_j^{(m)}}$,
\item divide the $j$th column of $H(D)$ by $D^{\tilde l_j^{(d)}}$ and multiply the same column by $D^{\tilde l_j^{(m)}}$.
\end{enumerate}
Then
\begin{enumerate}
\item the $j$th component of $\mbox{\boldmath $y$}_k$ becomes
\begin{equation}
y_k^{(j)}\rightarrow y_{k+l_j^{(d)}-l_j^{(m)}}^{(j)} ,
\end{equation}
\item the $j$th component of $\mbox{\boldmath $e$}_k$ becomes
\begin{equation}
e_k^{(j)}\rightarrow e_{k-\tilde l_j^{(d)}+\tilde l_j^{(m)}}^{(j)} .
\end{equation}
\end{enumerate}
After shifting $e_{k-\tilde l_j^{(d)}+\tilde l_j^{(m)}}^{(j)}$ by $l$ time units ($l$ is independent of $j$), compare the time-index of $e_{k+l-\tilde l_j^{(d)}+\tilde l_j^{(m)}}^{(j)}$ and that of $y_{k+l_j^{(d)}-l_j^{(m)}}^{(j)}$. If the two time-indices coincide, then $y_k^{(j)}$ and $e_k^{(j)}$ have ``relatively'' the identical shift. This condition is written as
\begin{equation}
l=(l_j^{(d)}+\tilde l_j^{(d)})-(l_j^{(m)}+\tilde l_j^{(m)})~(1\leq j \leq n) ,
\end{equation}
where $l$ is a constant independent of $j~(1\leq j \leq n)$. (In the following, this condition is denoted as ``$C_{SR}$''.) 

\subsection{Special cases}
{\it Case 1:} Only division is applied both to the columns of $G(D)$ and to those of $H(D)$.
\par
From the assumption, $l_j^{(m)}=\tilde l_j^{(m)}=0$. Hence, we have
\begin{equation}
l=l_j^{(d)}+\tilde l_j^{(d)} .
\end{equation}
Here, assume that either $l_j^{(d)}$ or $\tilde l_j^{(d)}$ is $0$. Define the sets $L_G$ and $L_H$ as
\begin{eqnarray}
L_G &\stackrel{\triangle}{=}& \{j:l_j^{(d)}=l\}=\{j:\tilde l_j^{(d)}=0\} \\
L_H &\stackrel{\triangle}{=}& \{j:\tilde l_j^{(d)}=l\}=\{j:l_j^{(d)}=0\} .
\end{eqnarray}
In words, $L_G$ is the set of columns of $G(D)$ from which $D^l$ is factoring out, whereas $L_H$ is the set of columns of $H(D)$ from which $D^l$ is factoring out. Note that $L_G$ and $L_H$ are disjoint and the relation
\begin{equation}
L_G\cup L_H=\{1, 2, \cdots, n\}
\end{equation}
holds. In the following, we call this kind of transformations ``type-1''.
\par
{\it Example 2:} Consider the relation
\begin{eqnarray}
G_2(D) &=& (D+D^2, D^2, 1+D) \nonumber \\
&\Leftrightarrow& H_2(D)=\left(
\begin{array}{ccc}
1& 0& D \\
D& 1+D& 0 
\end{array}
\right) .
\end{eqnarray}
Choosing $l=1$, $L_G=\{1, 2\}$, and $L_H=\{3\}$, we have
\begin{eqnarray}
G_2'(D) &=& (1+D, D, 1+D) \nonumber \\
&\Leftrightarrow& H_2'(D)=\left(
\begin{array}{ccc}
1& 0& 1 \\
D& 1+D& 0 
\end{array}
\right) .
\end{eqnarray}
\par
{\it Case 2:} Division and multiplication are separately applied either to the columns of $G(D)$ or to the columns of $H(D)$.
\par
Without loss of generality, assume that division is applied to the columns of $G(D)$, whereas multiplication is applied to the columns of $H(D)$. From the assumption, $l_j^{(m)}=\tilde l_j^{(d)}=0$. Hence, we have
\begin{equation}
l=l_j^{(d)}-\tilde l_j^{(m)} .
\end{equation}
In particular, set $l=0$. Then we have
\begin{equation}
l_j^{(d)}=\tilde l_j^{(m)}~(\stackrel{\triangle}{=}l_j) .
\end{equation}
This is equivalent to dividing the $j$th column of $G(D)$ by $D^{l_j}$ and multiplying the $j$th column of $H(D)$ by $D^{l_j}$. In the following, we call this kind of transformations ``type-2''.
\par
{\it Example 3:} Consider the relation
\begin{eqnarray}
G_3(D) &=& (1+D, 1, D+D^2) \nonumber \\
&\Leftrightarrow& H_3(D)=\left(
\begin{array}{ccc}
D& 0& 1 \\
1& 1+D& 0 
\end{array}
\right) .
\end{eqnarray}
Choosing $l_3^{(d)}=\tilde l_3^{(m)}=1$, we have
\begin{eqnarray}
G_3'(D) &=& (1+D, 1, 1+D) \nonumber \\
&\Leftrightarrow& H_3'(D)=\left(
\begin{array}{ccc}
D& 0& D \\
1& 1+D& 0 
\end{array}
\right) .
\end{eqnarray}
Note that $H_3'(D)$ can be reduced to
\begin{equation}
H_3''(D)=\left(
\begin{array}{ccc}
1& 0& 1 \\
1& 1+D& 0 
\end{array}
\right) .
\end{equation}
\par
Type-1 and type-2 transformations form a subclass of general transformations defined in Section III-A. However, these transformations are quite effective.

\subsection{Property of transformations}
Observe that in Example 2 and Example 3, the GH Relation is preserved after type-1 and type-2 transformations. It is shown that this property holds in general. Assume that the relation $G(D)\Leftrightarrow H(D)$ holds. Also, assume that a pair of transformations which satisfies the condition $C_{SR}$ is applied to $G(D)$ and $H(D)$. Let $G'(D)$ and $H'(D)$ be the resulting matrices, respectively. Then we have the following.
\newtheorem{pro}{Proposition}
\begin{pro}The relation $G'(D)\Leftrightarrow H'(D)$ holds.
\end{pro}
\begin{IEEEproof}Fix $p,~q~(1\leq p \leq n-m,~1\leq q \leq m)$ arbitrarily. Let
\begin{equation}
(g_{p1}(D), \cdots, g_{pj}(D), \cdots, g_{pn}(D))
\end{equation}
be the $p$th row of $G(D)$. Then the $(p,j)$ element of $G'(D)$ is given by
\begin{equation}
g_{pj}(D)\frac{D^{l_j^{(m)}}}{D^{l_j^{(d)}}} .
\end{equation}
Similarly, defining the $q$th row of $H(D)$ as
\begin{equation}
(h_{q1}(D), \cdots, h_{qj}(D), \cdots, h_{qn}(D)) ,
\end{equation}
the $(q,j)$ element of $H'(D)$ is given by
\begin{equation}
h_{qj}(D)\frac{D^{\tilde l_j^{(m)}}}{D^{\tilde l_j^{(d)}}} .
\end{equation}
Then the $(p,q)$ element $h'_{pq}$ of $G'(D)H'^T(D)$ is given by
\begin{eqnarray}
h'_{pq} &=& \sum_{j=1}^ng_{pj}(D)\frac{D^{l_j^{(m)}}}{D^{l_j^{(d)}}}h_{qj}(D)\frac{D^{\tilde l_j^{(m)}}}{D^{\tilde l_j^{(d)}}} \nonumber \\
&=& \sum_{j=1}^ng_{pj}(D)h_{qj}(D)D^{(l_j^{(m)}+\tilde l_j^{(m)})-(l_j^{(d)}+\tilde l_j^{(d)})} \nonumber \\
&=& \frac{1}{D^l}\sum_{j=1}^ng_{pj}(D)h_{qj}(D) .
\end{eqnarray}
Since $G(D)\Leftrightarrow H(D)$, $\sum_{j=1}^ng_{pj}(D)h_{qj}(D)=0$. Hence, we have $h'_{pq}=0$.
\end{IEEEproof}

\section{Simultaneous reduction of $G(D)$ and $H(D)$}
The discussion in the previous section implies that $G(D)$ and $H(D)$ can be reduced simultaneously, if reduction is possible. Assume that the relation $G(D)\Leftrightarrow H(D)$ holds. Let $\nu$ and $\nu^{\perp}$ be the overall constraint lengths of $G(D)$ and $H(D)$, respectively. If both $G(D)$ and $H(D)$ are {\it canonical} [4], [5], then we have $\nu=\nu^{\perp}$. Here, apply a pair of transformations which satisfies the condition $C_{SR}$ to $G(D)$ and $H(D)$. Denote by $\nu'$ and $\nu'^{\perp}$ the overall constraint lengths of the modified matrices $G'(D)$ and $H'(D)$, respectively. Note that the relation $G'(D)\Leftrightarrow H'(D)$ still holds from Proposition 1. Hence, if necessary, by modifying equivalently, we have $\nu'=\nu'^{\perp}$. Therefore, if the strict inequality $\nu'<\nu~(\nu'^{\perp}<\nu^{\perp})$ holds, then $G(D)$ and $H(D)$ are reduced simultaneously. That is, we have the following.
\begin{pro}Assume that the relation $G(D)\Leftrightarrow H(D)$ holds. Also, assume that a pair of transformations which satisfies the condition $C_{SR}$ is applied to $G(D)$ and $H(D)$. In this case, if $G(D)$ is reduced, then $H(D)$ is equally reduced, and vice versa.
\end{pro}
\par
{\it Example 4:} 
Assume that
\begin{eqnarray}
G_4(D) &=& (1+D+D^2, D, D^4+D^5) \nonumber \\
\Leftrightarrow H_4(D) &=& \left(
\begin{array}{ccc}
D^3& D^2& 1 \\
D& 1+D+D^2& 0 
\end{array}
\right) .
\end{eqnarray}
Note that both $G_4(D)$ and $H_4(D)$ are canonical and the equality $\nu=\nu^{\perp}=5$ holds. Choosing $l=1$, $L_G=\{2, 3\}$, and $L_H=\{1\}$, let us apply a type-1 transformation. Then we have
\begin{eqnarray}
G_4'(D) &=& (1+D+D^2, 1, D^3+D^4) \nonumber \\
\Leftrightarrow H_4'(D) &=& \left(
\begin{array}{ccc}
D^2& D^2& 1 \\
1& 1+D+D^2& 0 
\end{array}
\right) .
\end{eqnarray}
Also, let us apply a type-2 transformation with $l_3^{(d)}=\tilde l_3^{(m)}=2$. Then we have
\begin{eqnarray}
G_4''(D) &=& (1+D+D^2, 1, D+D^2) \nonumber \\
\Leftrightarrow H_4''(D) &=& \left(
\begin{array}{ccc}
D^2& D^2& D^2 \\
1& 1+D+D^2& 0 
\end{array}
\right) .
\end{eqnarray}
Since $H_4''(D)$ is reduced to
\begin{equation}
H_4'''(D)=\left(
\begin{array}{ccc}
1& 1& 1 \\
1& 1+D+D^2& 0
\end{array}
\right) ,
\end{equation}
we finally have
\begin{eqnarray}
G_4''(D) &=& (1+D+D^2, 1, D+D^2) \nonumber \\
\Leftrightarrow H_4'''(D) &=& \left(
\begin{array}{ccc}
1& 1& 1 \\
1& 1+D+D^2& 0 
\end{array}
\right) .
\end{eqnarray}
In this example, the overall constraint lengths are reduced from $\nu=\nu^{\perp}=5$ to $\nu'=\nu'^{\perp}=2$.
\par
{\it Remark:} The reduction process is not unique. In the above example, if a type-2 transformation is applied to $G_4(D)$ and $H_4(D)$ with $l_3^{(d)}=\tilde l_3^{(m)}=3$, then we have
\begin{eqnarray}
G_4^*(D) &=& (1+D+D^2, D, D+D^2) \nonumber \\
\Leftrightarrow H_4^*(D) &=& \left(
\begin{array}{ccc}
D^3& D^2& D^3 \\
D& 1+D+D^2& 0 
\end{array}
\right) \nonumber \\
\simeq H_4^{**}(D) &=& \left(
\begin{array}{ccc}
D& 1& D \\
D& 1+D+D^2& 0 
\end{array}
\right) ,
\end{eqnarray}
where ``$\simeq$'' means equivalent. Here, choosing $l=1$, $L_G=\{2\}$, and $L_H=\{1, 3\}$, let us apply a type-1 transformation. Then we have $G_4''(D)\Leftrightarrow H_4'''(D)$.

\section{Simultaneous code/error-trellis reduction}
Assume that the relation $G(D)\Leftrightarrow H(D)$ holds. Let $T_c$ be the code-trellis associated with $G(D)$. It is assumed that $T_c$ is terminated in the all-zero state at $t=N$. Denote by $T_e$ the corresponding error-trellis. Note that each code-path {\boldmath $y$} in $T_c$ corresponds to the unique error-path {\boldmath $e$} in $T_e$ by way of the received data {\boldmath $z$}. Here, apply a pair of transformations which satisfies the condition $C_{SR}$ to $G(D)$ and $H(D)$. (Let $G'(D)$ and $H'(D)$ be the resulting matrices.) Then from Proposition 2, it is reasonable to think that $T_c$ and $T_e$ are reduced simultaneously. In fact, we have the following.
\begin{pro}Assume that a pair of transformations which satisfies the condition $C_{SR}$ is applied to $G(D)$ and $H(D)$. In this case, if the code-trellis associated with $G(D)$ is reduced, then the error-trellis based on $H^T(D)$ is equally reduced, and vice versa.
\end{pro}
\begin{IEEEproof}Denote by $\mbox{\boldmath $e$}'$ the shifted version of {\boldmath $e$}. Assume that the set of shifted error-paths $\{\mbox{\boldmath $e$}'\}$ is represented using the reduced error-trellis $T_e'$ based on $H'^T(D)$. Note that there exists a one-to-one correspondence between the code-paths $\{\mbox{\boldmath $y$}\}$ and the error-paths $\{\mbox{\boldmath $e$}\}$. Also, from the assumption of the transformations, the identical shifts are generated both in the subsequences of a code-path {\boldmath $y$} and in those of the corresponding error-path {\boldmath $e$}. Hence, the set of shifted code-paths $\{\mbox{\boldmath $y$}'\}$ is also represented using the reduced code-trellis $T_c'$ associated with $G'(D)$. That is, if one trellis is reduced, then the other trellis is equally reduced.
\end{IEEEproof}

{\it Example 5:}
\begin{figure}[tb]
\begin{center}
\includegraphics[width=7.5cm,clip]{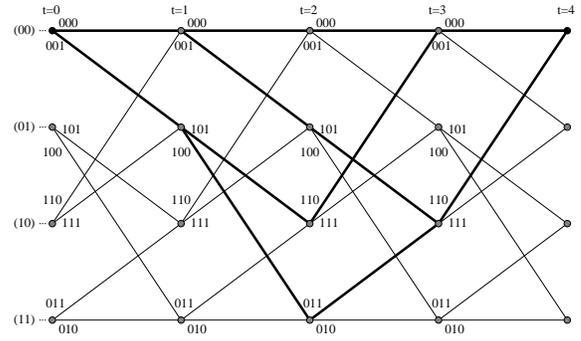}
\end{center}
\caption{Example code-trellis associated with $G_2(D)$.}
\label{Fig.1}
\end{figure}
\begin{figure}[tb]
\begin{center}
\includegraphics[width=8.0cm,clip]{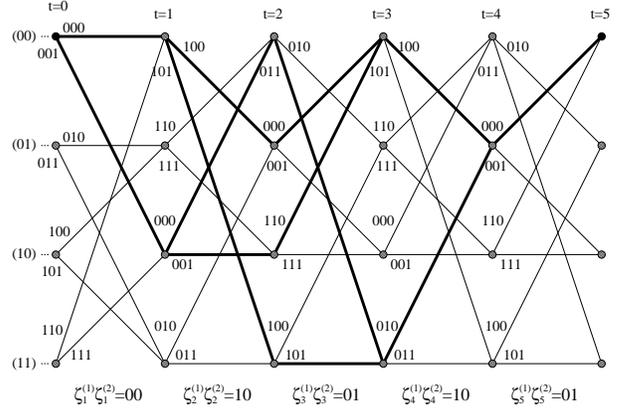}
\end{center}
\caption{Example error-trellis based on $H_2^T(D)$.}
\label{Fig.2}
\end{figure}
Consider the relation $G_2(D)\Leftrightarrow H_2(D)$. Fig.1 shows the code-trellis associated with $G_2(D)$. Note that the trellis is terminated in the all-zero state $(00)$ at $t=4$. The corresponding error-trellis based on $H_2^T(D)$ is shown in Fig.2. A received data {\boldmath $z$} is assumed to be
\begin{equation}
\mbox{\boldmath $z$}=\mbox{\boldmath $z$}_1~\mbox{\boldmath $z$}_2~\mbox{\boldmath $z$}_3~\mbox{\boldmath $z$}_4~\mbox{\boldmath $z$}_5=001~000~011~010~000 ,
\end{equation}
where $\mbox{\boldmath $z$}_5=000$ is the ``imaginary'' received data at $t=5$. The syndrome sequence is given as
\begin{equation}
\mbox{\boldmath $\zeta$}=\mbox{\boldmath $\zeta$}_1~\mbox{\boldmath $\zeta$}_2~\mbox{\boldmath $\zeta$}_3~\mbox{\boldmath $\zeta$}_4~\mbox{\boldmath $\zeta$}_5=00~10~01~10~01 .
\end{equation}
As we have already seen in Example 2, if the first and second components of $\mbox{\boldmath $y$}_k$ are shifted left by the unit time and if the third component of $\mbox{\boldmath $e$}_k$ is shifted right by the unit time, then $G_2(D)$ and $H_2(D)$ are reduced simultaneously. Denote by $G_2'(D)$ and $H_2'(D)$ the modified generator and parity-check matrices after transformation, respectively. The corresponding code and error-trellises are shown in Fig.3 and Fig.4, respectively.
\begin{figure}[tb]
\begin{center}
\includegraphics[width=8.0cm,clip]{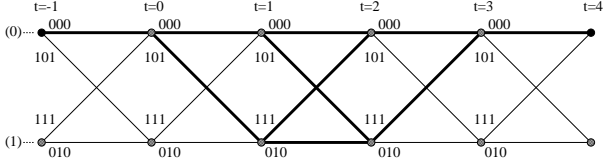}
\end{center}
\caption{Reduced code-trellis associated with $G_2'(D)$.}
\label{Fig.3}
\end{figure}
\begin{figure}[tb]
\begin{center}
\includegraphics[width=8.0cm,clip]{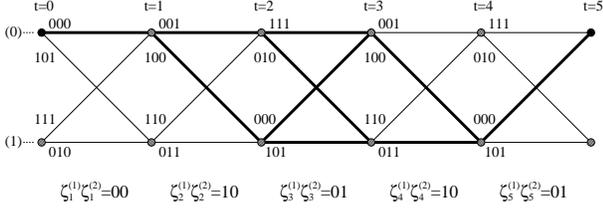}
\end{center}
\caption{Reduced error-trellis based on $H_2'^T(D)$.}
\label{Fig.4}
\end{figure}
\par
First, consider the reduced error-trellis in Fig.4. In this example, it is defined as $e_k'^{(3)}\stackrel{\triangle}{=}e_{<k-1>}^{(3)}$, where $<t>$ denotes $t\:\mbox{mod}\:5$. Since $\mbox{\boldmath $e$}_5=000$, we have $e_1'^{(3)}=e_{<0>}^{(3)}=e_5^{(3)}=0$ using the relation $e_k'^{(3)}=e_{<k-1>}^{(3)}$. That is, the third error-bit of the branch from $t=0$ to $t=1$ must be $0$. Similarly, the first two error-bits of the branch from $t=4$ to $t=5$ must be $00$. Then we have four admissible error-paths:
\begin{eqnarray}
\mbox{\boldmath $e$}_{p_1}' &=& 000~001~010~011~000 \nonumber \\
\mbox{\boldmath $e$}_{p_2}' &=& 000~001~111~100~000 \nonumber \\
\mbox{\boldmath $e$}_{p_3}' &=& 000~100~101~011~000 \nonumber \\
\mbox{\boldmath $e$}_{p_4}' &=& 000~100~000~100~000 . \nonumber
\end{eqnarray}
Here, noting the relation $e_k'^{(3)}=e_{<k-1>}^{(3)}$, we cyclically shift the third bit of each $\mbox{\boldmath $z$}_k$ to the right by the unit time and make the modified received data $\mbox{\boldmath $z$}'$ for $H_2'^T(D)$. $\mbox{\boldmath $z$}'$ is given by
\begin{eqnarray}
\mbox{\boldmath $z$}' &=& \mbox{\boldmath $z$}_1'~\mbox{\boldmath $z$}_2'~\mbox{\boldmath $z$}_3'~\mbox{\boldmath $z$}_4'~\mbox{\boldmath $z$}_5' \nonumber \\
&=& 000~001~010~011~000 .
\end{eqnarray}
Note that if $\mbox{\boldmath $z$}'$ is inputted to $H_2'^T(D)$, then the same syndrome sequence $\mbox{\boldmath $\zeta$}=00~10~01~10~01$ as for $H_2^T(D)$ is obtained.
\par
Next, consider the reduced code-trellis in Fig.3. Since $\mbox{\boldmath $y$}_0=000$, we have $y_4'^{(i)}=y_{<5>}^{(i)}=y_0^{(i)}=0~(i=1, 2)$. That is, the first two code-bits of the branch from $t=3$ to $t=4$ must be $00$. Similarly, the third code-bit of the branch from $t=-1$ to $t=0$ must be $0$. Here, to each of admissible error-paths in Fig.4, we add the modified received data $\mbox{\boldmath $z$}'$. Then we have
\begin{eqnarray}
\mbox{\boldmath $y$}_{p_1}' &=& 000~000~000~000~000 \nonumber \\
\mbox{\boldmath $y$}_{p_2}' &=& 000~000~101~111~000 \nonumber \\
\mbox{\boldmath $y$}_{p_3}' &=& 000~101~111~000~000 \nonumber \\
\mbox{\boldmath $y$}_{p_4}' &=& 000~101~010~111~000 . \nonumber
\end{eqnarray}
We observe that the obtained paths completely coincide with those in Fig.3. That is, the two trellises associated with $G_2(D)$ and $H_2^T(D)$ have been reduced simultaneously.

\section{Conclusion}
We have shown that the code-trellis and the error-trellis for a convolutional code can be reduced simultaneously. The proposed method is based on the fact that if the identical shifts occur both in the components of $\mbox{\boldmath $y$}_k$ and in the components of $\mbox{\boldmath $e$}_k$, then the two trellises are reduced simultaneously, if reduction is possible. We have obtained the general transformations which generate the identical shifts both in the subsequences of {\boldmath $y$} and in those of {\boldmath $e$}. We have shown that these transformations preserve the GH Relation. Using this property, we have shown that reduction of $G(D)$ and $H(D)$ is accomplished simultaneously, if it is possible. Moreover, we have shown that the corresponding two trellises are also reduced simultaneously. These results again imply that a code/error-trellis construction using shifted code/error-subsequences is very effective. We remark that a parity-check matrix with the form described in the paper appears in [11] in connection with a class of LDPC convolutional codes. We think [10] that the proposed method is useful for reducing the state complexity of the code/error-trellis for such an LDPC convolutional code.






%

\end{document}